\documentclass[twocolumn,preprintnumbers,secnumarabic,amsmath,amssymb,nofootinbib,floatfix]{revtex4}
\usepackage{varioref,exscale,latexsym,amsmath,amssymb}
\usepackage{graphicx}
\usepackage{epsfig}
\usepackage{slashbox}

\usepackage{graphicx}
\usepackage{dcolumn}
\usepackage{bm}

\def\beq{\begin{equation}}

\def\eeq{\end{equation}}

\def\beqa{\begin{eqnarray}}

\def\eeqa{\end{eqnarray}}

\begin{document}

\title{{\bf Self-healing of unitarity in effective field theories and the onset of new physics }}

\medskip\
\author{Ufuk Aydemir${}^{1}$}
\email[Email: ]{uaydemir@physics.umass.edu}
\author{Mohamed M. Anber${}^{2}$}
\email[Email: ]{manber@physics.utoronto.ca}
\author{ John F. Donoghue${}^{1}$}
\email[Email: ]{donoghue@physics.umass.edu}
\affiliation{${}^{1}$Department of Physics,
University of Massachusetts\\
Amherst, MA  01003, USA\\
${}^{2}$Department of Physics,
University of Toronto\\
Toronto, ON, M5S1A7, Canada\\
}

\begin{abstract}
In effective field theories it is common to identify the onset of new physics with the violation of tree-level unitarity. However, we show that this is parametrically incorrect in the case of chiral perturbation theory, and is probably theoretically incorrect in general. In the chiral theory, we explore perturbative unitarity violation as a function of the number of colors and the number of flavors, holding the scale of the``new physics'' (i.e. QCD) fixed. This demonstrates that the onset of new physics is parametrically uncorrelated with tree-unitarity violation. When the latter scale is lower than that of new physics, the effective theory must heal its unitarity violation itself, which is expected because the field theory satisfies the requirements of unitarity. In the chiral theory, the self-healing results in a resonant structure with scalar quantum numbers. In the electroweak variant of this argument, the structure must have the properties of the Higgs and must couple proportional to the mass in both gauge boson and fermion scattering. A similar example can be seen in the case of general relativity coupled to multiple matter fields, where iteration of the vacuum polarization diagram restores unitarity. We present arguments that suggest the correct identification should be connected to the onset of {\em inelasticity} rather than unitarity violation. We describe how the onset of inelasticity can occur in the effective theory, although it does not appear possible to predict the onset reliably.
\end{abstract}
\maketitle


\section{Introduction}
Effective field theories are low energy representations of more complete high energy theories. They contain the degrees of freedom that are active at low energy. However the effective theory generally lacks some of the degrees of freedom of the high energy theory - that is the ``new physics'' that becomes active when probed at a high enough energy.

Some effective field theories have a useful perturbative treatment at low energies, but that perturbative expansion runs into difficulties as the energy increases. When the perturbation involves powers of the energy, one of the difficulties is the violation of tree-level unitarity. The effective theory satisfies unitarity order by order in its perturbative energy expansion, and loop corrections restore tree-level unitarity but still leave higher order violations.  Calculationally this is similar to the way that renormalizeable theories satisfy unitarity order by order in the perturbative coupling constant expansion. However, a perturbative expansion in the energy always becomes difficult at high enough energy.

Does the violation of tree-level unitarity also signal the onset of the new physics of the high energy theory? Historically this argument dates back to work in the weak interactions by Lee, Quigg and Thacker \cite{Lee} who proposed the connection as a constraint on the mass of the Higgs. It has been elaborated on further in the literature \cite{furtherwork}. While this connection has some intuitive appeal, there has been no real proof. The problems of the perturbative expansion need not signal new physics, and it is at least conceivable that the effective theory could enjoy some range of non-perturbative validity before new physics is manifest.

Moreover, there are some theoretical ideas that suggest that the effective theory could be valid at {\em all} energies. One is the proposal for ``classicalization'' \cite{Dvali:2010jz} where classical solutions to the effective theory become important at high energy and solve the perturbative difficulties. For chiral theories these are soliton solutions and for general relativity, which also behaves as an effective field theory at low energy, the classical solutions are proposed to be black holes. Related ideas have been used in an attempt to understand gravity beyond the Planck scale \cite{Giddings:2011xs}. Another direction proposed to solve the perturbative problems is that of ``asymptotic safety'' \cite{Weinberg:1980gg, asymsafety} where the expansion in energy is tamed by the existence of an ultraviolet fixed point. While this was originally proposed for general relativity, it also comes in a chiral realization \cite{Percacci:2012mx}. If these ideas are correct there are no new degrees of freedom to be uncovered, but rather a non-perturbative realization of the effective theory.

In this paper we present evidence that the two phenomena of tree-unitarity violation and the onset of new physics are {\em separate} in practice in QCD-like theories of different numbers of colors and flavors. We show that they are parametrically disconnected. Moreover, for many values of these parameters, the unitarity violation happens in elastic scattering well below the energy where the QCD degrees of freedom are produced. In such a case, there is no other option but that the apparent unitarity violation must be solved within the effective field theory without recourse to the new degrees of freedom. We refer to this phenomenon as ``self-healing'' and we can see how it operates to the extent that perturbation theory can still be used.

Moreover, we will argue that {\em inelasticity} is more important than tree-unitarity. The unitarity arguments focus on the elastic $2\to 2$ scattering channel of Goldstone bosons. In essence, our reasoning is that in a theory like QCD where there is a duality between the quark-gluon description and the hadron-level description, the production of explicit quark or gluon degrees of freedom would not be dual to a state with only 2 Goldstone bosons, but would require more than these two hadronic degrees of freedom. In practice for real QCD, inelasticity does occur at a higher energy than the violation of tree-unitarity and appears more related to the opening up of the quark and gluon excitations. Moreover, in nonperturbative attempts such as classicalization, the classical solutions also require more than just two of the fundamental fields. We attempt to partially fill-out this argument in Sec. 6, but we find that the effective theory has limited predictive power for the onset of inelasticity.

\section{Parametric independence}

Field theories employ a unitary time development operator. The scattering matrix is derived from that operator
\begin{equation}
S_{ji} = {}_{out}<j| U(+\infty,-\infty) |i>_{in}\,,
\end{equation}
which then satisfies
\begin{equation}
S^\dagger S =S S^\dagger =1\,.
\end{equation}
Perturbative expansions of the S-matrix, whether in terms of a coupling constant or in terms of the energy, will then satisfy unitarity order by order in the expansion parameter.

For elastic scattering in the region below inelastic thresholds, there is only a single channel involved. Unitarity then implies
\begin{eqnarray}\label{unitarityeq}
\mbox{Im} T_{IJ}&=&\sigma |T_{IJ}|^2\underset{{\tiny \mbox{chiral limit}}}{\implies}\mbox{Im} T_{IJ}=|T_{IJ}|^2,\nonumber\\
\sigma&=&(1-\frac{4 m^2}{s})^{1/2}.
\label{impart}
\end{eqnarray}
Here the amplitude is labeled by the angular momentum $J$ and isospin $I$ (or other internal variables). As a consequence of Eq.(\ref{unitarityeq}), $T^{I}_{J}$ can be parameterized as
\begin{eqnarray}\label{TIJparametrization}
T_{IJ}=\frac{1}{\sigma}\,e^{i\delta_{{\tiny \mbox{IJ}}}}\,\mbox{sin}\delta_{IJ}.
\end{eqnarray}
From Eq. (\ref{TIJparametrization}), the basic consequence of unitarity is
\begin{eqnarray}\label{basicconsequenceofunitarity}
|\mbox{Re}\,T_{IJ}\,|\,\sigma\leq \frac{1}{2} \underset{{\tiny \mbox{chiral limit}}}{\implies} |\mbox{Re}\,T_{IJ}\,|\leq \frac{1}{2}.
\end{eqnarray}
This is the expression traditionally used in a test of tree-unitarity. The standard argument is that at the energy for which Eq. (\ref{basicconsequenceofunitarity}) is violated at tree-level one requires new physics to cure the unitarity violation.

In QCD with massless quarks\footnote{We will consistently ignore fermion masses as they are inessential to our argument.} the low energy limit of the theory is described by Goldstone bosons and the resulting effective field theory, chiral perturbation theory, has been well studied. For two massless fermions, the Goldstone particles are pions, and the strongest scattering occurs in the scalar-isoscalar channel with amplitude\cite{weinberg}
\begin{equation}
T_{00}^{\mbox{\scriptsize tree}} = \frac{s}{16\pi F^2}\,,
\end{equation}
where $F$ is the pion decay constant in the chiral limit $F=86.5$~MeV. This amplitude violates the tree unitarity condition at $434$~MeV.

QCD can also be considered as a function of the number of colors, $N_c$. For large $N_c$, keeping $g^2N_c$ fixed keeps the overall scale of QCD fixed. In this limit, hadron masses and thresholds stay the same, while scattering amplitudes vanish as $1/N_c$. In practice many large-$N_c$ counting rules seem to work well even at $N_c=3$. In the following, we will use the leading order $N_c$ counting rules and ignore sub-leading corrections. This is sufficient for our goal of showing the different parametric dependence of unitarity violation relative to new physics. For us the important scaling relation, holding the scale of QCD fixed, is that $F \sim \sqrt{N_c}$, which then tells us that $T_{00}^{\mbox{\scriptsize tree}} \sim 1/N_c$. Using this scaling rule, the energy of the violation of tree-unitarity scales like $E\sim  \sqrt{N_c}$. The numerical dependence on the number of colors can be seen in Table I.

\begin{table}[bh]
\centerline{
\begin{tabular}{|c|ccccc|}
  \hline
  \backslashbox{$N_f$}{$N_c$} & 2& 3 & 4 & 5 & 6  \\\hline
    2 & 354  &434   & 501  & 560 & 613  \\
    3 & 289  &  354 & 409 & 457 & 501   \\
    4 & 250  & 307  & 354 & 396 & 434  \\
    5 &  224 & 274  & 317 & 354 & 388  \\
    6 &  204 &  250 & 289 & 323 & 354 \\
  \hline
\end{tabular}
}
\caption{The table shows the energy (in MeV) at which tree unitarity is violated, varying $N_f$ and $N_c$ while holding the scale of QCD fixed to leading order in the parameters.}
\label{Nfdependence}
\end{table}

QCD can also be considered as a function of the number of light flavors. In the real world, the strange quark is not dynamically active at the scale where pion scattering violated tree-unitarity. However, chiral perturbation theory readily allows the analysis of the limit where the strange quark is massless. With a little more effort, one could consider a world where the charmed quark or the other heavy quarks were to become massless, and indeed scattering amplitudes have been analyzed for the case of $N_f$ light flavors \cite{Chivukula:1992gi, Bijnens:2011fm}. The leading effect for our problem is the identification of the channel with the strongest scattering, which then is the one that violates unitarity at the lowest energy. The generalization of the isospin singlet pi-pi channel is the flavor singlet meson-meson channel. This amplitude has a linear dependence on the number of flavors
\begin{equation}
T_{00}^{N_f} = \frac{N_f s}{32\pi F^2}\,,
\end{equation}
and the factor of $N_f$ is a combinatoric in origin. There can be sub-leading dependence on the number of light flavors through their influence on $F$ and the scale of QCD. For example, the masses of light hadrons have a small dependence on the number of light flavors which could be analyzed through the quark contribution to the QCD beta function and the trace anomaly. Likewise the shift in $F$ due to taking the strange quark to be massless can be analyzed. However, again the leading parametric dependence is sufficient for our purposes, and we will not include sub-leading dependence. The resulting energies corresponding to the violation of tree unitarity can be seen in the $N_f$ dependence of Table I.

The results of Table I provide evidence that the location of the violation of tree-unitarity is parametrically disconnected from the scale of QCD. This location cannot be used to predict the onset of new QCD physics beyond the effective field theory.

Moreover, the energies listed in Table I are below any reasonable measure of the onset of new physics. The quark and gluon structure of QCD - the ``new physics'' - becomes potentially visible around 1.5-2 GeV. Historically this was the range of energies where the first evidence of jettiness in $e^+e^-$ reactions was uncovered. At around this energy also, the resonances become numerous enough to provide a hint of the quark substructure. The lowest energy providing evidence beyond the effective field theory was at the rho meson, 0.77 GeV, but this by itself did not give enough information on the nature of the new physics of QCD\footnote{Also note that the rho meson does not solve the unitarity problem.}. Whatever the precise definition of the onset of new physics, it is clear that there are situations in QCD-like theories where the unitarity problem must be solved within the effective theory before new physics arises.

\section{Self-healing in chiral theories}

It is straightforward to provide a unitary representation of the chiral predictions. For a start, the tree level amplitude not only predicts the leading real part of the amplitude, but also the leading imaginary part
\begin{equation}
 \mbox{Im} T_{00}^{leading} = \left[T_{00}^{\mbox{\scriptsize tree}}\right]^2 =\left[\frac{N_f s}{32\pi F^2}\right]^2\,.
 \label{TreeIm}
\end{equation}
These real and imaginary parts can be arranged in an amplitude
\begin{equation}
T_{00} = \frac{T_{00}^{\mbox{\scriptsize tree}}}{1 -i T_{00}^{\mbox{\scriptsize tree}}}\,,
\label{treeiteration}
\end{equation}
which satisfies elastic unitarity exactly at all energies. A geometric expansion of the denominator indicates that this comes from an iteration of multiple rescatterings in the elastic channel. Also note that for the physical $\pi\pi$ $N_c=3$ situation, this produces a pole on the second sheet at the energy
\begin{eqnarray}
\sqrt{s}=(434-i\,434)\,\mbox{MeV}\,.
\end{eqnarray}
This compares favorably with the best determination of the sigma pole \cite{caprini}, using data plus chiral constraints and the Roy equations, of
\begin{eqnarray}
\sqrt{s}=(441-i\,272)\,\mbox{MeV}\,.
\end{eqnarray}
If we were to allow $N_c$ and $N_f$ to vary, the corresponding mass and width following from the tree amplitude is always equal to the location of tree unitarity violation listed in Table I. Within the range of the parameters considered there, the $\sigma$ pole could be as light as 204~MeV.

\begin{figure*}
\centering
\begin{tabular}{cc}
\epsfig {file=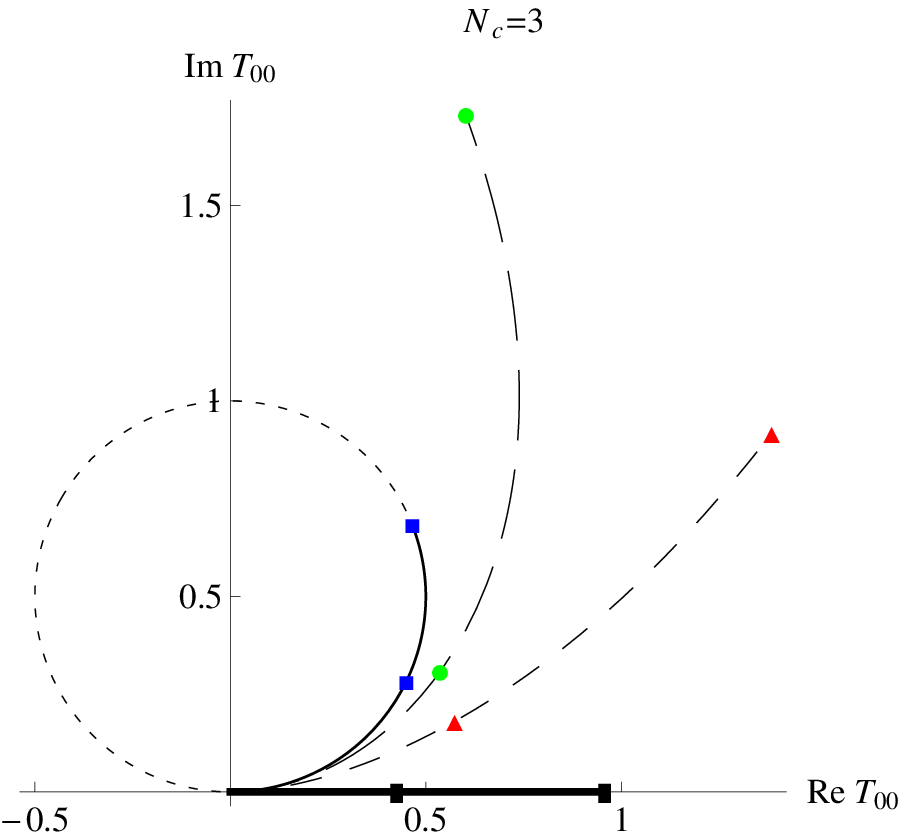,width=0.5\linewidth,clip=}&
\epsfig {file=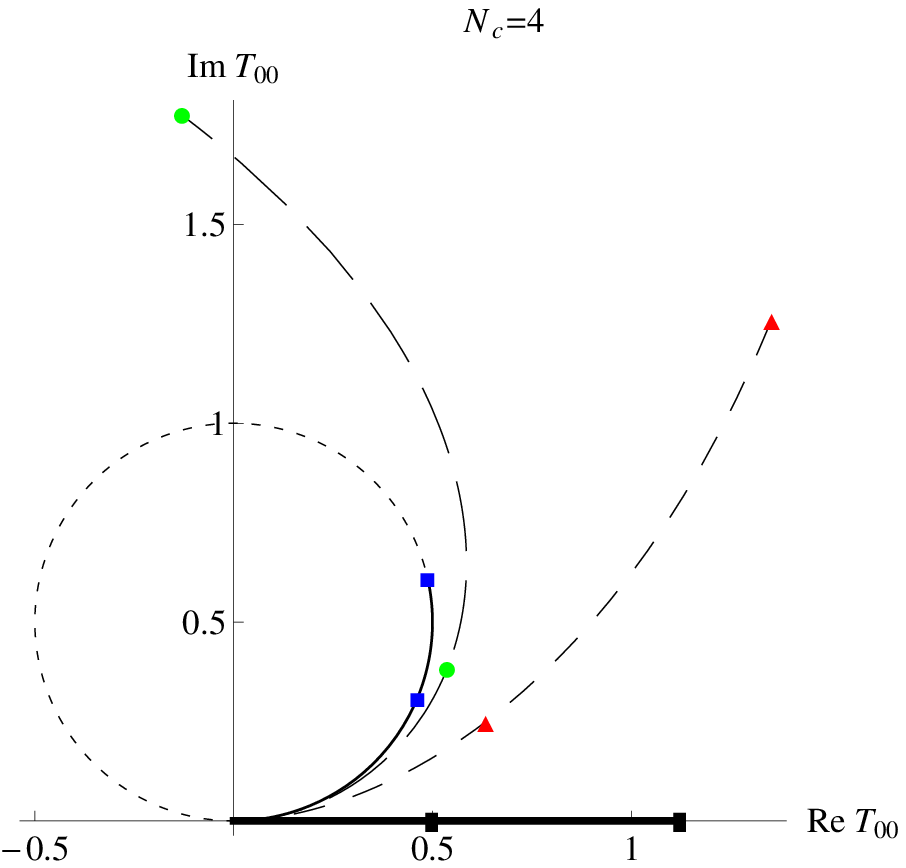,width=0.5\linewidth,clip=} \\\\
\epsfig {file=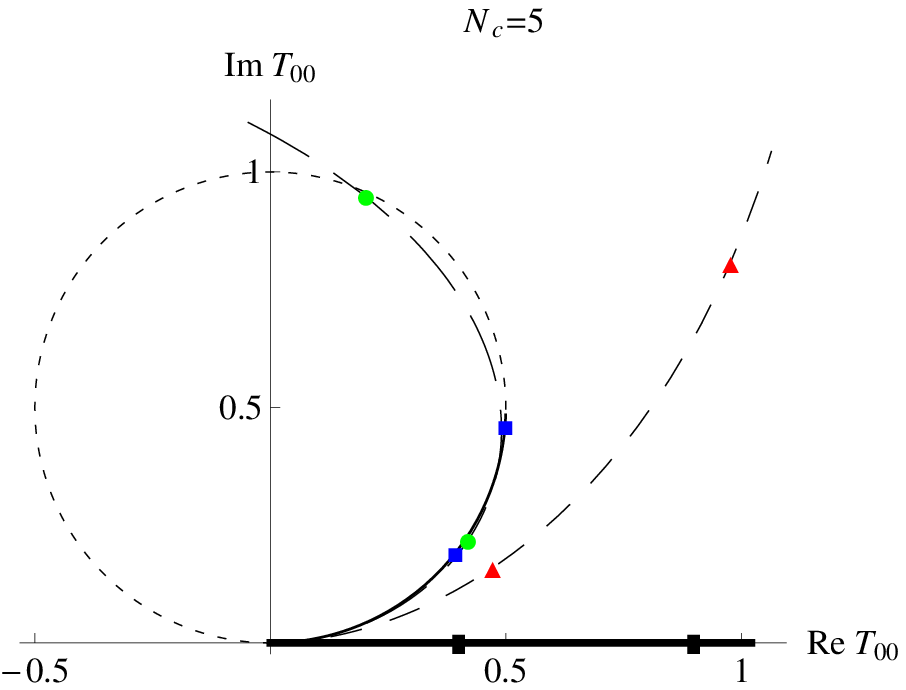,width=0.5\linewidth,clip=}&
\epsfig {file=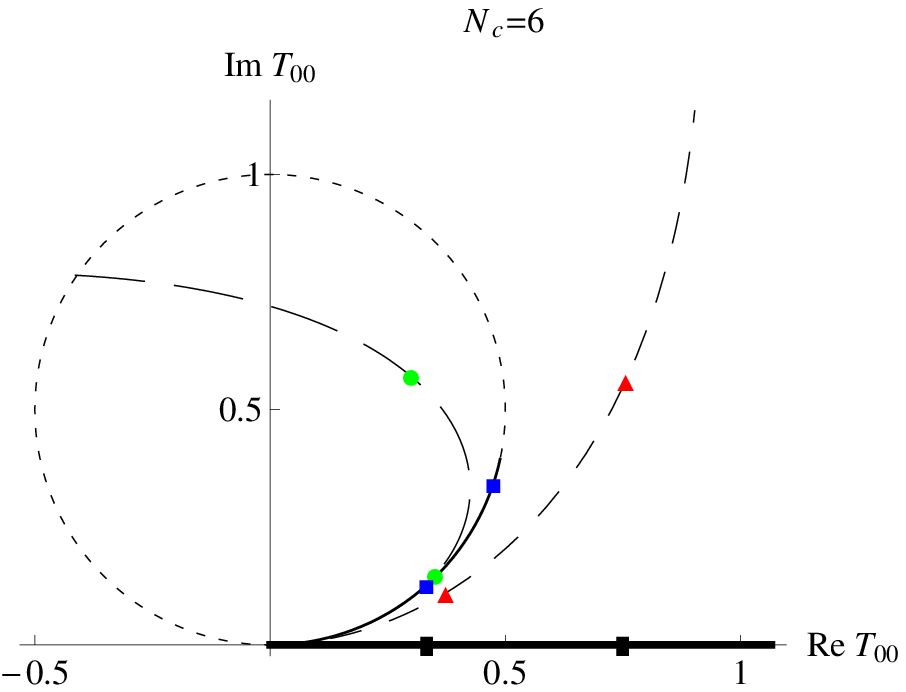,width=0.5\linewidth,clip=}
\end{tabular}
\caption{Argand plots for SU(2) amplitudes. Dotted line is unitary circle. The (thick) solid, dashed, and long-dashed lines represent tree-level, NLO, and NNLO ChPT results, respectively. The thin solid line is NNLO IAM amplitude. The first and second points on the plots represent $\sqrt{s}=400$ MeV and $\sqrt{s}=600$ MeV in the first graph and $\sqrt{s}=500$ MeV and $\sqrt{s}=750$ MeV in the others, respectively. These points are denoted as  rectangulars, triangles, circles, and squares on the plots of tree-level, NLO, NNLO, and IAM amplitudes, respectively. We observe that $N_c=3$ and $N_c=4$ amplitudes diverge significantly (more than 10 $\%$) from unitary circle at very low energies ($\approx 150$ MeV and $\approx 190$ MeV. However, their behavior strongly implies self-healing at higher orders. $N_c=5$ and $N_c=6$ amplitudes satisfy elastic unitarity up to $\approx765$ MeV and $\approx600$ MeV. In the case of $N_c=5$, we see some deviation starting at $\approx 600$ MeV (toward inner part of the circle), but this remains insignificant.}
\label{argandplotsSU2}
\end{figure*}

This procedure can be systematically improved using chiral perturbation theory. The chiral framework for $\pi\pi$ scattering is well known \cite{weinberg}-\cite{Bijnens:1999hw}, culminating in the work of Colangelo, Gasser and Leutwyler (CGL)\cite{cgl}. The generalization for $N_f$ flavors has been worked out by Bijnens and Lu \cite{Bijnens:2011fm}. We will discuss the approach to unitarity implied in these works. The simplest method for obtaining exact unitarity is the Inverse Amplitude Method(IAM) \cite{IAM}, and even at NLO this provides a quite reasonable parameterization of the results of CGL up to the inelastic threshold. The IAM has also been carefully explored as a function of the number of colors \cite{pelaez}. We will use this and also extend the method as a function of $N_f$.

First consider the physical case, $N_f=2$, $N_c=3$. Here the scattering results are known to order $p^6$. In FIG. \ref{argandplotsSU2} we show the convergence of the scalar-isoscalar amplitude towards the unitarity circle. The tree level result of course runs along the real axis. The energy dependence is marked on this axis. This is corrected at one-loop (order $p^4$) to include the imaginary part of Eq. \ref{TreeIm}, plus corrections to the real part. It starts the approach to the unitarity circle. The approach is enhanced at two-loop order. Because the amplitudes are expanded in powers of the energy, the imaginary part of the amplitude always satisfies the unitarity relation of Eq. \ref{impart} with both sided of the equation being evaluated at the same power of the energy. Specifically, when the imaginary part is treated to order $p^4$, the unitarity relation is satisfied to the extent that the real part is treated to order $p^2$. For the $p^6$ treatment of the imaginary part, the result correctly follows from the $p^2$ and $p^4$ portions of the real part of the amplitude. However, at each order there are higher order terms which are not exactly predicted. Chiral perturbation theory satisfies unitarity order by order in the energy expansion.

\begin{figure}[ht]
\centerline{
\includegraphics[width=12cm]{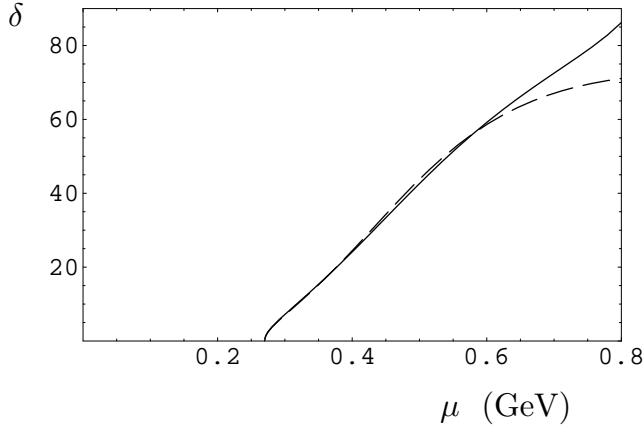}
}
\vspace*{-0.2in}

 \caption{The scalar-isoscalar $\pi\pi$ phase shift from CGL (solid line) and that
 found from the inverse amplitude method (dashed line) at energy $E=\mu$. The agreement is excellent up
 to about 700 MeV.}
 \label{phasecgl}
\end{figure}

\begin{figure}[ht]
\centerline{
 \includegraphics[width=12cm]{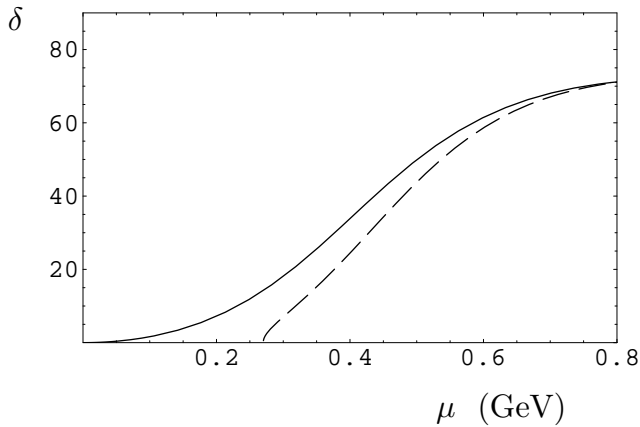}
}
\vspace*{-0.2in}

 \caption{The scalar-isoscalar $\pi\pi$ phase shifts using the IAM in the physical
 case (dashed line) and in the chiral limit (solid line) at energy $E=\mu$.}
 \label{chiral phases}
\end{figure}

The full result converges to the unitarity circle. A fully unitarity result can be obtained by using the inverse amplitude method. A dispersion theory justification is given in Ref. \cite{IAM} but in practice the method amounts to using chiral perturbation theory to predict the inverse amplitude rather than the amplitude itself. For the real world, including masses, the comparison with the phase shift from the IAM (at order $p^2$) to the CGL analysis of data is shown in FIG. \ref{phasecgl}. We can use this result to go to the chiral limit, with a phase shift shown in Fig. \ref{chiral phases}. This is then our approximation for the amplitude on the unitarity circle.

\begin{figure*}
\centering
\begin{tabular}{cc}
\epsfig {file=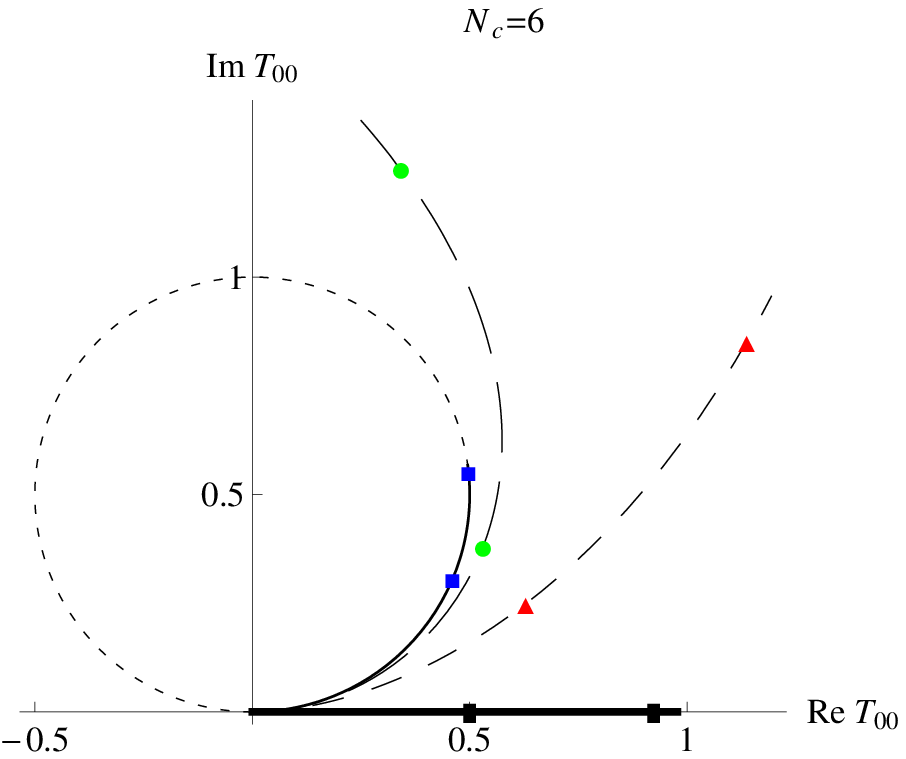,width=0.5\linewidth,clip=}&
\epsfig {file=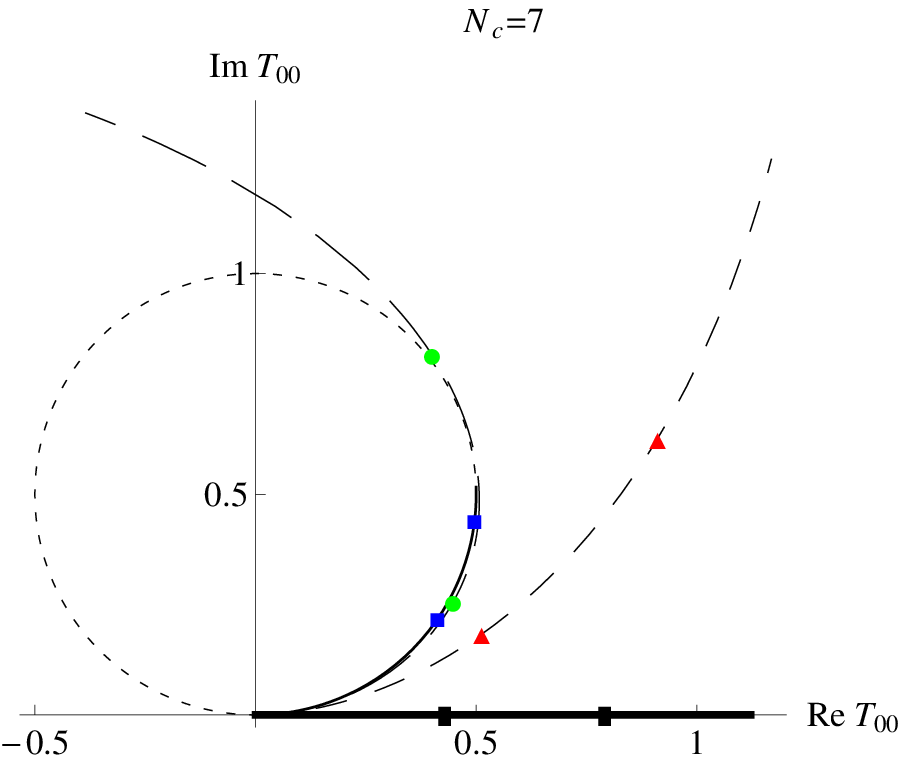,width=0.5\linewidth,clip=}
\end{tabular}
\caption{Argand plots for SU(3) amplitudes. The line and symbol representations are same in Fig. \ref{argandplotsSU2}, except the first and second points in these plots represent $\sqrt{s}=400$ MeV and $\sqrt{s}=680$ MeV where the latter value is the point beyond which elastic unitarity (for NNLO ChPT) is violated in the second graph.}
\label{argandplotsSU3}
\end{figure*}

Using these ingredients, one can see that this convergence to unitarity works also at other values of $N_c$ and $N_f$. Some of these other cases are shown in the figures. The $N_c$ dependence of the chiral coefficients at order $p^4$ are well known. For those at $p^6$ we scale with factors of $N_c^2$ for triple trace lagrangians, $N_c$ for double trace ones and keep single trace ones with a constant coefficient. For $N_f$ scaling we use the results of Bijnens and Lu\cite{Bijnens:2011fm}, with results shown in Fig.\ref{argandplotsSU3} . These also illustrate the nature of the convergence of the energy expansion.

Within the inverse amplitude method we can then also explore the evolution of the pole on the second sheet with loop order, again as a function of the number of colors and flavors. The results are shown in Tables II, III and IV. We see that this location is modified by higher order corrections. However, there remains always a pole, which is connected to the unitarity of the full amplitude.

\begin{table}
\centerline{
\begin{tabular}{|l|lll|lll|}
  \hline
 $ $ & \multicolumn{3}{c}{$m_{\sigma}$ (MeV)} & \multicolumn{3}{c}{Unit. viol. en. (MeV) } \\ \hline
  $N_c$ & tree-lev. & 1-loop & 2-loop & tree-lev. & 1-loop & 2-loop \\ \hline
   2 & 354 & 283  & 258  & 354  & 298   &  298 \\
   3 & 434 & 303 & 287 & 434 & 376 & 385 \\
   4 & 501 & 291 & 292 & 501 & 447 &474 \\
   5 & 560 & 250 & 278 & 560 & 515 & 845 \\
   6 & 613 & 173 & 247 & 613 & 583 & 900 \\
  \hline
\end{tabular}
}
\caption{$N_c$ dependence in SU(2) of pole mass ($m_{\sigma}$) and the energy at which the minimal consequence of unitarity is violated.}
\label{NcdependenceSU2}
\end{table}

\begin{table}
\centerline{
\begin{tabular}{|l|lll|lll|}
  \hline
 $ $ & \multicolumn{3}{c}{$m_{\sigma}$ (MeV)} & \multicolumn{3}{c}{Unit. viol. en. (MeV) } \\ \hline
  $N_c$ & tree-lev. & 1-loop & 2-loop & tree-lev. & 1-loop & 2-loop \\ \hline
   2 & 289  &  245  &  220  & 289   & 238  & 234  \\
   3 & 354  & 283  & 258  & 354   & 298  & 298  \\
   4 & 409  &  299 & 279  & 409 & 351  & 357 \\
   5 & 457  &  299 & 289  & 457  & 401  & 415  \\
   6 & 501  & 285  & 288  & 501  & 448  & 477  \\
  \hline
\end{tabular}
}
\caption{$N_c$ dependence in SU(3) of pole mass ($m_{\sigma}$) and the energy at which the minimal consequence of unitarity is violated.}
\label{NcdependenceSU3}.
\end{table}

\begin{table}[bh]
\centerline{
\begin{tabular}{|l|lll|lll|}
  \hline
 $ $ & \multicolumn{3}{c}{$m_{\sigma}$ (MeV)} & \multicolumn{3}{c}{Unit. viol. en. (MeV) } \\ \hline
  $N_f$ & tree-lev. & 1-loop & 2-loop & tree-lev. & 1-loop & 2-loop \\ \hline
   2 & 434  & 303  & 287  & 434   & 376  & 385  \\
   3 & 354  &  283 & 258  & 354 & 298  & 298 \\
   4 & 307  &  261  &  235  & 307  & 252 & 248  \\
   5 & 274  & 237  & 214   & 274  & 222 & 216   \\
   15 & 158  & 130  & 122  & 158  & 121 & 115  \\
   80  & 69   & 50  & 50   & 69  &  50  & 47  \\
   100 & 61  & 44  & 44    & 61 &  44  &  41  \\
  \hline
\end{tabular}
}
\caption{$N_f$ dependence of pole mass ($m_{\sigma}$) and the energy at which the minimal consequence of unitarity is violated. The color number is fixed to $N_c=3$.}
\label{Nfdependence}.
\end{table}

\section{The electroweak effective theory and the inevitability of a Higgs}

There is a Goldstone sector associated with electroweak symmetry breaking. Within the Standard Model this is associated with the Higgs sector, but even in the absence of the standard Higgs field, the interactions longitudinal gauge bosons obey the constraints of an SU(2) effective lagrangian. The original argument about tree-unitarity violation\cite{Lee} followed from the observation that the Higgs sector provides a solution to the unitarity violation. This is certainly the correct solution when the Higgs is light - one has a complete weakly coupled perturbative theory. Electroweak radiative corrections and recent LHC data seem to favor a light Higgs, so this is quite possibly the solution favored by nature.

However, because the tree-unitarity argument has traditionally been used to point to new physics at the TeV scale, we will make some brief comments about what our results suggest if one tries to push the Higgs to very large mass or remove it entirely from the theory. In this case the effective lagrangian description takes over \cite{Dobado:1997jx}. The scattering of longitudinal gauge bosons is equivalent to the meson-meson scattering discussed previously. Tree unitarity is violated at about 1.2 TeV. However, this does not mean that there is visible new physics at this scale. Dobado et al \cite{Dobado:1999xb} have discussed the consequences in a manner that is close to our treatment above. Pushing new physics to high energy corresponds to taking values of the coefficients of the chiral lagrangian to be no larger than the size implied by naive dimensional analysis.\footnote{Dimensional analysis says that the coefficients cannot be much smaller than the size that arises from the scale dependence of loop corrections, which for the weak interaction implies a relative size of $(4\pi v)^2$.} Dobado et al show that the unitarized amplitudes essentially always has a scalar pole, analogous to the sigma discussed above. This plays the role of the Higgs in the strongly coupled theory.

The remarkable feature is that this scalar pole also shows up in $f \bar{f}$ scattering with a mass proportional to the fermion mass, again like the Higgs. This follows as an extension of the work of Appelquist and Chanowitz \cite{Appelquist:1987cf}, who showed that there was tree unitarity violation in the scattering of $f\bar{f} \to W_L W_L$ which is proportional to the quark mass. However, to the extent that only the longitudinal gauge sector is involved, the same physics that unitarizes the gauge boson channel also unitarizes the scattering starting from the $q\bar{q}$ initial state. This can be treated by Omnes function \cite{Omnes:1958hv}, which was invoked in a related context in Ref. \cite{dgl}. (See also Ref. \cite{dhgamma} for the application in photon scattering.) The Omnes function describes strong elastic rescattering and is calculated from the elastic scattering phase shift $\delta (s)$
\begin{equation}
\Omega (s) = \exp\left[\frac{s}{\pi} \int ds' \frac{\delta (s')}{s(s' -s +i\epsilon) }\right]\,.
\end{equation}
The scattering amplitude then has the structure
\begin{equation}
a_0 (f\bar{f}\to W^+_L W^-_L)= \frac{\sqrt{2}G_F\sqrt{s}}{16\pi}m_f\Omega (s)
\end{equation}
proportional to the fermion mass. The same rescattering phases enter both reactions. Higher order fermion loops will modify this construction, but the result will continue to satisfy unitarity.

We note however, that as a practical matter, it is not simple to identify the scalar pole in these unitarized amplitudes because the result can be a wide pole far from the real axis, which need not show up as a conventional bump in the cross section. This is known from explicit studies \cite{Dobado:1999xb} but the conclusion is apparent from the 40 years of work that was needed to unravel the $\sigma$ in low energy QCD \cite{cgl}. Perhaps it is fortunate that present indications favor a light Higgs.

\section{Gravity coupled to multiple fields}

In Ref. \cite{han}, Han and Willenbrock considered gravity coupled up to many light fields. In a procedure analogous to the treatment of large numbers of flavors in Sec. 2, one identifies the linear combination of scalars, fermions and vectors that scatters elastically in the $J=2$ channel, and calculate the resulting tree amplitude, in this case in the $J=2$ channel because it proceeds through s-channel graviton exchange. The result is
\begin{equation}
T_2^{tree}= - \frac{G_N N s}{40}\,,
\end{equation}
where $N$ is given in terms of the number of light real scalars, fermions and vectors, $N = N_s/3 +N_f +4N_v$. This violates tree unitarity at
$s= \frac{20}{G_N N}$. The authors interpret this violation quite strictly, claiming that new physics must enter before $\sqrt{s} = 6 \times 10^{18}$~GeV in the Standard Model, $\sqrt{s} = 4.6 \times 10^{18}$~GeV in the MSSM, and correspondingly lower if one lets $N$ become arbitrarily large.

However, given our experience with QCD-like theories above, we disagree with this interpretation. The evidence of chiral theories with larger numbers of flavors is that the onset of new physics does not track the $N_f$ dependence of the violation of tree-unitarity. The effective theory heals its unitarity issue at energies between the unitarity point and the beginning of new physics.

\begin{figure}[ht]
\centerline{
\includegraphics[width=.25\textwidth]{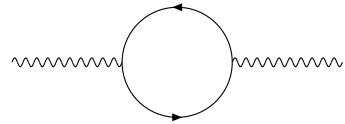}
}
\caption{Vacuum polarization graph. }
\label{VP}
\end{figure}

Gravity coupled to light fields is an effective field theory whose formulation respects unitarity perturbatively. The theory also has the ability to heal its unitarity violation. Indeed, the authors of Ref. \cite{han} identify the mechanism for this self-healing. The required imaginary part to the amplitude at next order comes from the cut in the vacuum polarization diagram (see FIG. \ref{VP}). Summing the tree plus the iterations of vacuum polarization loop yields a result that satisfies elastic unitarity exactly
\begin{equation}
\label{exact elastic unitarity}
T_2 = \frac{T_2^{\mbox{\scriptsize tree}}}{1 - \frac{\mbox{Re} T_2^{\mbox{\scriptsize 1-loop}}}{T_2^{\mbox{\scriptsize tree}}} -i T_2^{\mbox{\scriptsize tree}}}\,.
\end{equation}
The real part of the one loop amplitude requires renormalization, and this is accomplished as usual in the gravitational effective field theory \cite{jfd}.

This healing mechanism can be demonstrated explicitly. For simplicity, let us consider scalars coupled to gravity:
\begin{eqnarray}
\nonumber
S=\int d^4 x\sqrt{-g}\left[\frac{R}{16\pi G_N}+\frac{1}{2}\left(g^{\mu\nu}\partial_\mu \phi\partial_\nu \phi+\lambda R\phi^2\right)\right]\,.\\
\end{eqnarray}
Two of us have elsewhere considered the scattering amplitude of the reaction $A_i+A_i\rightarrow A_j+A_j$, where $ i \ne j$, for minimally coupled scalars, $\lambda=0$,  through s-channel graviton exchange along with one loop gravitational corrections \cite{runningG}. Here, we consider the case of non-minimally coupled scalars, specifically we note that  $\lambda=-1/6$ corresponds to conformally coupled scalars. The s-channel tree amplitude of $A_i+A_i\rightarrow A_j+A_j$ is given by
\begin{eqnarray}
\label{tree A gravity}
{\cal A}_{\mbox{\scriptsize tree}}=\frac{8\pi G_N}{s}\left[2s^2\lambda(3\lambda+1)+ut \right]\,.
\end{eqnarray}
Now, we can perform partial wave expansion in angular momentum $J$
\begin{eqnarray}
{\cal A}(s,\cos\theta)=32\pi \sum_{J}(2J+1)T_J(s)P_J(\cos\theta)\,,
\end{eqnarray}
and, as usual, parametrize the $t$ and $u$ channel respectively as $t=-s(1-\cos\theta)/2$ and $u=-s(1+\cos\theta)/2$. Hence, we find for the tree-level partial wave $T_2^{\mbox{\scriptsize tree}}$
\begin{eqnarray}
T_2^{\mbox{\scriptsize tree}}=-\frac{G_NN_s s}{120},
\end{eqnarray}
which is independent of $\lambda$.

 Next, we turn to the one loop amplitude of $A_i+A_i\rightarrow A_j+A_j$. Using only the vacuum polarization diagram that contains scalars (we exclude gravitons) we find
\begin{eqnarray}
\label{1loop gravity}
{\cal A}^{\mbox{\scriptsize 1-loop}}=-\frac{G_N^2N_s}{15}\left[s^2{\cal F}(\lambda)-tu \right]\log(-s)\,,
\end{eqnarray}
where ${\cal F}(\lambda)=1+20\lambda+180\lambda^2+720\lambda^3+1080\lambda^4$. Performing partial wave analysis for ${\cal A}^{\mbox{\scriptsize 1-loop}}$ we obtain
\begin{eqnarray}
\nonumber
\mbox{Re}T_{2}^{\mbox{\scriptsize 1-loop}}&=&-\frac{1}{\pi}\left(\frac{G_NN_ss}{120}\right)^2\log s\,,\\
\mbox{Im}T_{2}^{\mbox{\scriptsize 1-loop}}&=&\left(\frac{G_NN_ss}{120}\right)^2\,.
\end{eqnarray}
Hence, we have shown explicitly $|T_2^{\mbox{\scriptsize tree}}|^2=\mbox{Im}T_{2}^{\mbox{\scriptsize 1-loop}}$, and thus the elastic unitarity is achieved order by order. Summing the tree plus iterations of the one-loop yields as in Eq. \ref{exact elastic unitarity}
\begin{eqnarray}
\label{iteration partial wave}
T_2=\frac{-\left(\frac{G_NN_ss}{120}\right)}{1-\frac{1}{\pi}\left(\frac{G_NN_ss}{120}\right)\log(-s)}\,,
\end{eqnarray}
which satisfies the elastic unitarity exactly $\mbox{Im}T_2=|T_2|^2$.

Instead of iterating the partial waves, we instead can iterate the vacuum polarization graphs. The graviton self-energy of $N_s$ scalars circulating in the loop reads
\begin{eqnarray}
\nonumber
\Pi^{\alpha\beta,\mu\nu}(q)&=&-\frac{N_sG_N}{240\pi}\log\left(\frac{-q^2}{\mu^2}\right)\left[-6q^2(1+40\lambda/3\right.\\
\nonumber
&&\left.+40\lambda^2)\left(q^\alpha q^\beta g^{\mu\nu}+q^\mu q^\nu g^{\alpha\beta}\right)\right.\\
\nonumber
&&\left.-q^2\left(q^\alpha q^\mu g^{\beta\nu}+q^\alpha q^\nu g^{\beta\mu}+q^\beta q^\mu g^{\alpha\nu}\right.\right.\\
\nonumber
&&\left.\left. +q^\beta q^\nu g^{\alpha\mu}  \right)+q^4\left( g^{\alpha\nu}g^{\beta\mu}+g^{\alpha\mu}g^{\beta\nu}\right)\right.\\
\nonumber
&&\left.+6q^4(1+40\lambda/3+40\lambda^2)g^{\alpha\beta}g^{\mu\nu}\right.\\
&&\left.+8(1+10\lambda+30\lambda^2)q^\alpha q^\beta q^\mu q^\nu\right].
\end{eqnarray}
The polarization tensor can be written in the form
\begin{eqnarray}
\nonumber
\Pi^{\alpha\beta,\mu\nu}(q)&=&q^4A(q^2)L^{\alpha\beta}(q)L^{\mu\nu}(q)\\
\nonumber
&&-q^2B(q^2)\left[L^{\alpha\mu}(q)L^{\beta\nu}(q)+L^{\alpha\nu}(q)L^{\beta\mu}(q)\right.\\
&&\left. -2L^{\alpha\beta}(q)L^{\mu\nu}(q)\right],
\end{eqnarray}
where $L^{\mu\nu}(q)=\eta^{\mu\nu}-q^{\mu}q^{\nu}/q^2$.
Contracting $\Pi^{\alpha\beta,\mu\nu}(q)$ with $L^{\alpha\beta}(q)L^{\mu\nu}(q)$, and $L^{\alpha\mu}(q)L^{\beta\nu}(q)$ we obtain two equations in $A(q^2)$, and $B(q^2)$
\begin{eqnarray}
\nonumber
A(q^2)&=&-\frac{1}{30\pi}N_sG_N(1+10\lambda+30\lambda^2)\log\left(\frac{-q^2}{\mu^2}\right)\,,\\
B(q^2)&=&\frac{1}{240\pi}N_sG_Nq^2\log\left(\frac{-q^2}{\mu^2}\right)\,.
\end{eqnarray}
Here $\mu$ is related to the renormalization of the $R^2$ and $R_{\mu\nu}R^{\mu\nu}$ operators that appear at next order in the energy expansion.

The bare propagator takes the general form
\begin{eqnarray}
i{\cal D}^{\alpha\beta,\mu\nu}(q^2)=\frac{i}{2q^2}\left[L^{\alpha\mu}L^{\beta\nu}+L^{\alpha\nu}L^{\beta\mu}-L^{\alpha\beta}L^{\mu\nu} \right]\,,
\end{eqnarray}
while the quantum corrected propagator reads
\begin{eqnarray}
\nonumber
i{\cal D}'^{\alpha\beta,\mu\nu}&=&i{\cal D}^{\alpha\beta,\mu\nu}+i{\cal D}^{\alpha\beta,\gamma\delta}i\Pi_{\gamma\delta,\rho\tau}i{\cal D}^{\rho\tau,\mu\nu}\\
\nonumber
&=&\frac{i}{2q^2}(1+2B(q^2)) \left[L^{\alpha\mu}L^{\beta\nu}+L^{\alpha\nu}L^{\beta\mu}\right.\\
\nonumber
&&\left.-L^{\alpha\beta}L^{\mu\nu} \right]\\
&&-i\frac{A(q^2)}{4}L^{\alpha\beta}L^{\mu\nu}\,.
\label{dressed}
\end{eqnarray}
The first term above is the dressed propagator of $J=2$ graviton, while the second term is the contribution from the off-shell $J=0$ graviton. The second term drops out in the calculations of the matrix elements of conformally coupled scalar since in this case $T^\mu_\mu=0$.

In the large $N_s$ limit, keeping $G_NN_s$ small, we can sum up an infinite series of one-loop diagrams. Thus, the spin-2 part reads
\begin{eqnarray}
i{\cal D}^{\alpha\beta,\mu\nu}(q^2)=\frac{i\left[L^{\alpha\mu}L^{\beta\nu}+L^{\alpha\nu}L^{\beta\mu}-L^{\alpha\beta}L^{\mu\nu} \right]}{2q^2\left(1-\frac{N_sG_Nq^2}{120\pi}\log\left(-\frac{q^2}{\mu^2}\right)  \right)}\,.
\end{eqnarray}
The scattering amplitude $A+A \rightarrow B+B$ in the large $N_s$ limit can be read directly from the above dressed propagator
\begin{eqnarray}
\label{dressed A gravity}
{\cal A}^{\mbox{\scriptsize dressed}}=\frac{8\pi G_N}{s}\frac{2s^2\lambda(3\lambda+1)+ut}{1-\frac{G_NN_ss}{120\pi}\log\left(\frac{-s}{\mu^2}\right)}\,.
\end{eqnarray}
The difference between the $\lambda$ dependence in Eq. \ref{1loop gravity} and Eq. \ref{dressed A gravity} is due to the appearance of the $A(q^2)$ term in the propagator. In the conformal case, the difference vanishes and the expression of ${\cal A}^{\mbox{\scriptsize dressed}}$ agrees with Eq. \ref{1loop gravity}. However, this difference appears only in the $J=0$ channel, and hence does not alter the value of the partial amplitude $T_2$, which is solely due to the exchange of the spin-$2$ graviton.
Then, one can immediately obtain the partial wave amplitude $T_2$
\begin{eqnarray}
T_2=\frac{-\left(\frac{G_NN_ss}{120}\right)}{1-\frac{1}{\pi}\left(\frac{G_NN_ss}{120}\right)\log\left(\frac{-s}{\mu^2}\right)}\,,
\end{eqnarray}
which is identical to Eq. \ref{iteration partial wave}.

\section{Inelasticity and new physics}

The unitarity relation that we have studied above involves simple elastic scattering. It is doubtful that elastic scattering by itself can reveal the thresholds involving new degrees of freedom. By definition, the new degrees of freedom are different from the low energy particles. Something new must be produced. In a theory like QCD, we are used to considering a duality between the quark and hadron degrees of freedom. However, this is clearly not a one-to-one correspondence. A quark jet contains a large number of hadrons, mainly pions. The threshold for producing quark and gluon jets would then correspond to a large number of pions, not just the two which are involved in elastic scattering. The new physics must then be above the inelastic threshold and would be more associated with inelastic production than with the restoration of unitarity.

In this section we study the extent that the effective chiral theory could predict a threshold of significant inelasticity\footnote{In the real world, inelasticity in pi-pi scattering is associated with the opening of the $K-\bar{K}$ threshold. However, this is not in general the important inelasticity, which would correspond rather to multiple pions.}. With massless particles, there is no threshold for producing extra particles - there will be some inelasticitiy at any energy. However, there is a phase space penalty for inelastic channels - the phase space grows as a power of $s^{n-2}$, where $n$ is the number of particles produced. In particular we find that n-body phase space for massless particles behaves as
\begin{eqnarray}
\Pi_n(s)=\frac{1}{4}\frac{s^{n-2}}{(2\pi)^{2n-3}}\frac{1}{4^{n-2}\,\,(n-1)!(n-2)!}\,.
\end{eqnarray}
The growth as $s^{n-2}$ is clear simply from dimensional arguments. The combinatoric factors in the denominator suppress the production of multiple particles. However, in principle extra combinitoric factors can occur when taking matrix elements, so that it appears that a study of chiral lagrangians might have the potential to predict the energy at which the multiple pions becomes important.

However there is a variation of the large $N_c$ counting arguments that explains why the effective lagrangian is parametrically incapable of predicting the onset of large amounts of inelasticity. The coefficients in the chiral lagrangian have a maximum power of $N_c^1$, which is realized for single trace terms\cite{Gasser:1984gg}. The connection between the trace over the flavor indices and quark loops implies that multiple trace operators are suppressed by powers of $1/N_c$. When treating a reaction $\pi\pi \to n\pi$ , terms in the expansion of the lagrangian containing $n+2$ fields then brings out a factor of $1/F^{n+2}$. Since $F\sim \sqrt{N_c}$, the low energy amplitude for the production of $n$ pions behaves as $N_c^{-(1+n/2)}$. This implies that for any combination of chiral lagrangians the location where the $n$ particle reaction becomes important varies with $N_c$ in the limit where the scale of QCD is held fixed. The lagrangian cannot predict the inelastic threshold because of this different parametric behavior.

How then does QCD manage to keep a constant inelastic threshold in the large $N_c$ limit. The answer lies in the resonant production of radially excited mesons in the large $N_c$ limit \cite{terning}. High mass radially excited mesons can fall apart into jets of quarks and can provide a dual description to the direct jet production. Resonant production can overcome the large $N_c$ suppression. Consider for example the standard Breit-Wigner shape
\begin{equation}
\sigma \sim \frac{\Gamma^2}{(s-m^2)^2 + m^2\Gamma^2}\,.
\end{equation}
If $\Gamma$ is suppressed by a factor $1/N_c$, then when one is far from the pole the contribution of this resonance to scattering or to a chiral
Lagrangian will also be suppressed. However, if one is at the energy of the pole, the contribution is of order unity. For finite $N_c$ the resonance widths can overlap given the close spacing in mass and provide a representation of the continuum which is dual to the $q\bar{q}$ description. However, the chiral Lagrangian cannot reliably locate the location of the poles because the coefficients of the Lagrangian only feel the contribution of this resonance for $s$ close to zero energy.

\section{Discussion}

We have explored cases where the violation of tree unitarity is parametrically disconnected from the onset of new physics. Moreover, in cases where tree unitarity is violated well below the scale of new physics, we have seen how the unitarity problem can readily heal itself. In the gravitational example, the iteration of the vacuum polarization creates an unitary amplitude. The chiral case is similar, with an iteration of the basic scattering amplitude restoring unitarity. This resolution is not surprising because the effective field theories are obtained from the perturbative expansion of an S matrix that is unitary by construction. However, these results do indicate that tree-unitarity violation is {\em not} a good indicator of the onset of new physics.

We have argued that the most interesting scale is that where inelasticities become important. It is here where the new physics would emerge or the original theory would have a different phase. There is no requirement that a low energy effective theory be able to accurately predict the scale of the next new physics. We have looked at inelasticity in the effective chiral theory and argued that it is also parametrically not capable of predicting the onset of inelasticity.

These arguments do not change the general expectation that we expect that there will be new physics {\em somewhere} in the neighborhood of the energy where the effective theory becomes so strongly coupled as to be difficult to treat in perturbation theory. However, the desired onset of new physics is not predicted by the effective theory with any certainty nor control, and we have seen cases where there can be an order of magnitude difference. Indeed if it becomes possible to understand effective field theories beyond their perturbative region, it may be possible to push the need for new physics arbitrarily high in energy.

\section*{Acknowledgements}

The work of U.A. and J.D. has been supported in part by the NSF grant PHY - 0855119, and in part by the Foundational Questions Institute. The work of M.A. is supported by NSERC Discovery Grant of Canada.

\newpage
\providecommand{\href}[2]{#2}\begingroup\raggedright
\renewcommand\refname{References}

\end{document}